\documentclass[aps,pra,showpacs,floatfix,superscriptaddress,twocolumn]{revtex4-1}
\usepackage{graphicx}
\usepackage{dcolumn}
\usepackage{amsmath}
\usepackage{amssymb}
\usepackage{dsfont}
\usepackage{bm}
\usepackage{epstopdf}

\begin{document}

\title{Strongly trapped two-dimensional quantum walks}

\author{B. Koll\'ar}
\affiliation{Wigner RCP, SZFKI, Konkoly-Thege M. u. 29-33, H-1121 Budapest, Hungary}

\author{T. Kiss}
\affiliation{Wigner RCP, SZFKI, Konkoly-Thege M. u. 29-33, H-1121 Budapest, Hungary}

\author{I. Jex}
\affiliation{Department of Physics, Faculty of Nuclear Sciences and Physical Engineering, Czech Technical University in Prague, B\v
rehov\'a 7, 115 19 Praha 1 - Star\'e M\v{e}sto, Czech Republic}


\date{\today}

\begin{abstract}
Discrete time quantum walks (DTQWs) are nontrivial generalizations of random walks with a broad scope of applications. In particular, they can be used as computational primitives, and they are suitable tools for simulating other quantum systems. DTQWs usually spread ballistically due to their quantumness. In some cases, however, they can remain localized at their initial state (trapping). The trapping and other fundamental properties of DTQWs are determined by the choice of the coin operator. We introduce and analyze an up to now uncharted type of walks driven by a coin class leading to strong trapping, complementing the known list of walks. This class of walks exhibit a number of exciting properties with the possible applications ranging from light pulse trapping in a medium to topological effects and quantum search.
\end{abstract}

\maketitle

Quantum walks are analogs of classical random walks, gaining considerable attention since their introduction in the nineties \cite{Aharonov1993,Meyer1996}. They did prove themselves as interesting constructs which find their applications gradually \cite{Search,Entanglement,Sanders2003,Childs2009,Kendon2007,Shikano2010,Paparo2012,Leung2010,Kollar2012,Cedzich2013,Kollar2014} (for review see \cite{Reviews}). Recent experiments demonstrated their basic properties \cite{exp}, and stimulated further theoretical studies. While the early experiments have been limited to DTQWs on a line, recently, two-dimensional quantum walks have been successfully realized \cite{2DExp}.  Although a two-dimensional lattice is still considered simple in physics, two-dimensional DTQWs \cite{Mackay2002} display a rich variety of interesting effects. In particular, they exhibit a partial quantum speedup in search \cite{Search}.

The fundamental behavior of DTQWs is governed by the choice of the unitary operator acting on the internal structure of the particle --- the so-called coin operator. Similarly to random walks, DTQWs can be classified based on their spreading properties.  One of the classifications is based on the definition of the P\'olya number \cite{Polya1921}, dividing walks into recurrent and transient walks. Classically, this property depends only on the dimension of the underlying lattice. However, the quantum extensions of the P\'olya number \cite{Stefanak2008,Grunbaum2013,SinkoviczPreprint} exhibit significant differences. Apart from the dimension of the underlying lattice, the classification also depends on the choice of the coin operator and the internal state of an initially localized walker. In the family of two-dimensional recurrent quantum walks a rather interesting class can be found, namely that exhibiting the effect of trapping (localization) \cite{Mackay2002}. In this class the probability of finding the particle at its initial location throughout the whole evolution is non-vanishing. For leaving walks, i.e. when the walker is forced to leave its actual position during a single step, this effect does not have a natural classical counterpart, thus, it is a purely quantum interference phenomenon. One can post the natural question: Which coins exhibit trapping? First examples of such a coin were found in \cite{Inui2004,Watabe2008}.

In this paper we construct a large class of trapping coins, and point out important physical consequences for the corresponding DTQWs. In particular, we report the new effect of ``strong trapping". In addition to contributing to the understanding and classification of quantum walks, we comment on the implications --- possible application for the trapping of light, its relation to spatial quantum search algorithms, appearance of topological phases, and its spectral similarity to the so-called split step QWs. 

Let us begin with the formal definition of two-dimensional DTQWs, briefly. The composite Hilbert space of a DTQW is given as: $\mathcal{H} = \mathcal{H}_P \otimes \mathcal{H}_C$. In the present paper the position space $\mathcal{H}_P$ corresponds to a two-dimensional Cartesian lattice. We denote position Hilbert vectors with $ | \mathbf{x}  \rangle_P $, where $\mathbf{x} = (x,y)$ represents the position on the lattice. The coin space $\mathcal{H}_C$ is spanned by vectors corresponding to the directions: $| L \rangle, | D \rangle, | U \rangle, | R \rangle$. Throughout the present paper we expand operators on the coin space using this basis. We note that this coin Hilbert space can be viewed as a two-qubit Hilbert space with $ | L \rangle = | 00 \rangle, | D \rangle = | 01 \rangle, | U \rangle = | 10 \rangle, | R \rangle = |11 \rangle$. The time evolution is unitary: $U = S \cdot ( I_P \otimes C)$. Here, $S$ is the step operator responsible for the conditional displacement: $ S | \mathbf{x}  \rangle_P \otimes | c \rangle = | \mathbf{x} \oplus c \rangle_P \otimes | c \rangle $, where $c \in \{L,D,U,R\}$ corresponds to the directions and $\mathbf{x} \oplus c$ denotes the nearest neighbor of position $\mathbf{x}$ in the direction $c$.  By $C \in U(4)$ we denote the coin operator acting on the internal coin degree of freedom. Due to the translational invariance of the DTQWs we study, the global phase of the $U(4)$ coins is ignored. 
The time evolution in the quasi-momentum picture is given by
\begin{equation}
 \tilde{U} = \tilde{D} \cdot C\,,
 \label{time_evolution_fourier}
\end{equation}
where $\tilde{D}$ is the Fourier-transformed step operator $S$, i.e.
$\tilde{D} = \text{Diag} ( e^{-i k}, e^{-i l}, e^{i l}, e^{i k})$.
By $k,l$ we denote the quasi-momenta of the walk. The phase $\omega$ of a given $\lambda = e^{i \omega}$ eigenvalue of $\tilde{U}$ is frequently termed as quasi-energy.

Undoubtedly, the most notorious two-dimensional DTQW is the one driven by the Grover diffusion operator:
\begin{equation}
C^{(G)}_{i,j} = \frac{1}{2} - \delta_{i,j}\,,
\label{grover_coin}
\end{equation}
where $\delta_{i,j}$ is Kronecker's delta symbol.
Apart from its use in quantum walk based search algorithms \cite{Search}, the Grover walk is well-known for its trapping (localization) property \cite{Mackay2002,Inui2004}: The probability of finding the walker at its initial position at any time is non-vanishing for almost all walkers started at the origin. For the well chosen localized initial state
\begin{equation}
| \psi \rangle_{\text{init}}^{\text{NT}} = | \mathbf{x_0} \rangle_P \otimes \frac{1}{2} (| L \rangle - | D \rangle - | U \rangle + | R \rangle)
\label{gro_escape}
\end{equation}
the trapping is avoided and the walker escapes. A one-parameter generalization of the Grover walk exists  \cite{Watabe2008}, sharing the above listed traits:  state-dependent trapping and a single localized escaping-state. The extensive research dedicated to the Grover trapping also led to other interesting effects, namely dynamics exhibiting full revivals \cite{Stefanak2010}, a way to design recurrent quantum walks in arbitrary dimensions \cite{Stefanak2008}, and trapping in walks on percolation graphs \cite{Kollar2014NJP}.

In general, the trapping in DTQWs is caused by the presence of flat bands in the quasi-energy spectrum of the walk \cite{Inui2004}, which naturally depends on the choice of the coin operator. It was shown that the spectra of all trapping two-dimensional DTQWs contain at least two such flat bands with a $\pi$ difference between their quasi-energies \cite{Stefanak2010}. Finding trapping coins involves solving Eq. \eqref{time_evolution_fourier} for a pair of constant eigenvalues. However, it turns out that the direct solution in the complete $U(4)$ space is very demanding both numerically and analytically.

\begin{figure}[t!]
\begin{center}
\includegraphics[width=0.35\textwidth]{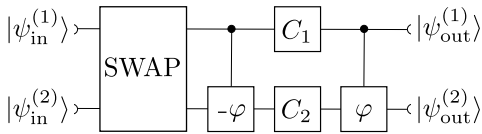} 
\end{center}
\caption{
Quantum circuit representing the coin class $\mathcal{C}$ defined in Eq. (\ref{Cclass}).
The abstract two-qubit space of $| \psi_{\text{in(out)}}^{(1)} \rangle$ and $| \psi_{\text{in(out)}}^{(2)} \rangle$ is spanned by basis vectors $ \{ | 00 \rangle, | 01 \rangle, | 10 \rangle, |11 \rangle \}$ correspond to
the four dimensional coin space of the two-dimensional quantum walk with $ | L \rangle = | 00 \rangle, | D \rangle = | 01 \rangle, | U \rangle = | 10 \rangle, | R \rangle = |11 \rangle$.
} \label{Figure2qubit_circ}
\end{figure}

Let us present a broad class of trapping coin operators using a constructive approach.
To begin, we define two arbitrary $SU(2)$ operators $U_1$ and $U_2$, thus, $\text{det}( U_{1(2)} )= 1$.
It is straightforward to see that the products $U_1 U_2$ and $U_2 U_1$ share the same pair of eigenvalues (they are unitary equivalent), and since $\text{det}( U_1 U_2) = \text{det}( U_2 U_1) = 1$ these eigenvalues are complex conjugate pairs.
Consequently, the following operator
\begin{equation}
\mathcal{\overline{U}}^2 = (U_2 \otimes U_1) \cdot  (U_1 \otimes U_2)\,,
\end{equation}
always has a constant eigenvalue $\mu = 1$ with the multiplicity of two.
We apply a unitary transformation keeping the constant eigenvalues:
\begin{equation}
\mathcal{U}^2 \equiv W \mathcal{\overline{U}}^2 W = W (U_2 \otimes U_1)  W W (U_1 \otimes U_2) W\,.
\label{eq_inprogress}
\end{equation}
Here, $W$ is the unitary  SWAP gate on a two-qubit system (coin space): $W \equiv | L \rangle\langle L | + | D \rangle\langle U | + | U \rangle\langle D | + | R \rangle\langle R |\,$.
Note that $W = W^{-1} = W^{\dagger}$, and we also inserted the identity $I = WW$ into Eq. (\ref{eq_inprogress}). Performing the SWAP operation on the first part of Eq. \eqref{eq_inprogress}, and taking a square root results in
$ \mathcal{U} = ( U_1 \otimes U_2 ) W $.
We substitute the arbitrary $SU(2)$ operators $U_{1(2)}$ with 
$ U_{1(2)} = \text{Diag}(e^{-i (k +(-) l)/2}, e^{i (k +(-) l)/2}) \cdot C_{1(2)} $, where $C_{k=\{1,2\}} \in SU(2)$, to arrive at the formula:
\begin{equation}
 \mathcal{U} = \tilde{D} (C_1 \otimes C_2) W = \tilde{U}\,,
\end{equation}
which is a time evolution operation of a two-dimensional DTQW [cf. Eq. (\ref{time_evolution_fourier})].
By its construction, $\mathcal{U}$ has at least two constant eigenvalues $\pm1$. These eigenvalues appear as flat bands at $\omega = 0,\pi$ quasi-energies in the spectrum. Thus, we constructed a trapping walk with the new coin class:
\begin{equation}
 \overline{\mathcal{C}} = \left(C_1 \otimes C_2 \right) W\,.
\end{equation}
One can observe that any operation commuting with $\tilde{D}$ and $W$ can generalize the coin $\overline{\mathcal{C}}$  without losing its trapping property (constant eigenvalues --- flat bands). Since $C_1 \otimes C_2$ cover all local single-qubit rotations, a non-separable operation is desirable to avoid redundancy. We found that the most suitable operation is the controlled phase gate: $P(\varphi) = \text{Diag}(1,1,1,e^{i \varphi})$. Thus, the general form of the trapping coin class ($\mathcal{C}$-class) is
\begin{equation}
 \mathcal{C} = P(\varphi) \left(C_1 \otimes C_2 \right) P(-\varphi) W\,,
 \label{Cclass}
\end{equation}
which contains $7$ real parameters ($+1$ for the global phase).
This coin class can be realized as the two-qubit quantum circuit illustrated in Fig. \ref{Figure2qubit_circ}.
We note that the $\mathcal{C}$-class contains the Grover coin [cf. Eq. \eqref{grover_coin}] and its known one-parameter generalization \cite{Watabe2008} as special cases.

Let us show some basic properties of the $\mathcal{C}$-class.
We define the elements of  $C_{k=\{1,2\}} \in SU(2)$:
\begin{equation}
C_{k} = \left( \begin{array}{rr}
e^{-i \alpha_{k}} \cos \delta_{k} & -e^{-i \beta_{k}} \sin \delta_{k} \\
e^{i \beta_{k}} \sin \delta_{k} & e^{i \alpha_{k}} \cos \delta_{k} 
\end{array} \right)\,.
\label{U2parametrisation}
\end{equation}
One can see that the time evolution operator $\tilde{U}$ of Eq. (\ref{time_evolution_fourier}) using the $\mathcal{C}$-class has two constant eigenvalues: $1$ and $-1$.
Solving the eigenvector problem, an inverse Fourier-transform reveals the stationary eigenstates corresponding to the eigenvalues $\lambda = \pm1$:
\begin{eqnarray}
| \psi (x,y)_{\lambda = \pm1} \rangle = & &  \nonumber \\
\frac{1}{2}\Big\{  | x, y \rangle_P & \otimes & ( -e^{-i \beta_1} \sin \delta_1  | L \rangle   - e^{-i \alpha_1} \cos \delta_1 | D \rangle)  \nonumber\\
\pm  | x, y+1 \rangle_P & \otimes & ( -e^{-i \beta_2} \sin \delta_2  | L \rangle  -  e^{-i \alpha_2} \cos \delta_2 | U \rangle)  \nonumber\\
\pm  | x+1, y \rangle_P & \otimes  & ( e^{i \alpha_2} \cos \delta_2  | D \rangle  -  e^{i (\beta_2 + \varphi)} \sin \delta_2 | R \rangle)  \nonumber\\
+  | x+1, y+1 \rangle_P & \otimes  & ( e^{i \alpha_1} \cos \delta_1  | U \rangle  -  e^{i (\beta_1 + \varphi)} \sin \delta_1 | R \rangle) \Big\} \,.  \nonumber\\
\label{localized_eigenstate}
\end{eqnarray}
An equally weighted superposition of two such eigenstates living on the same lattice sites $\frac{1}{\sqrt{2}} \left\{| \psi (x,y)_{\lambda = +1} \rangle  \pm | \psi (x,y)_{\lambda = -1} \rangle \right\}$ gives full reviving states. These states generalize the results of \cite{Stefanak2010} for the $\mathcal{C}$-class.

\begin{figure}[t!]
\begin{center}
\includegraphics[width=0.29\textwidth]{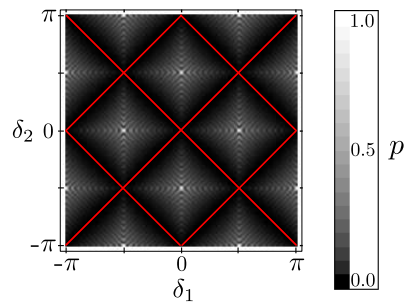}
\end{center}
\caption{ (Color online)
Minimum probability of finding the particle at its initial position after $40$ steps of a $\mathcal{C}$-class coin [see Eq.  (\ref{Cclass})] driven quantum walk. The probabilities are determined by numerically searching the initial states yielding the minimum probability. The coin is investigated in the $\delta_1, \delta_2$ parameter space with $ \alpha_1 = \alpha_2 = \beta_1 = \beta_2 = \frac{\pi}{2}$ and $\varphi=\pi$. The red lines correspond to the condition of Eq. (\ref{condition_of_superlocalization}) where strong trapping disappears. On these lines the minimum probability is naturally zero due to the appearance of the escaping state of Eq. (\ref{escapingstate}). However, where strong trapping is present, the minimum probability is larger than zero.
} \label{Fig_strtrap}\end{figure}

An initial state having a nonzero overlap with any of the eigenstates in Eq. (\ref{localized_eigenstate}) is trapped. We found that for most of the coins in the $\mathcal{C}$-class, all initially localized states are trapped. We term this novel phenomena as ``strong trapping". However, when the the coin parameters satisfy 
\begin{equation}
\cos 2 \delta_1 = \cos 2 \delta_2\,,
\label{condition_of_superlocalization}
\end{equation}
similarly to the Grover walk, a single initially localized state will lead to escape. For $\delta_2 = \delta_1 + k \cdot \pi \,,\, k \in \mathds{N^+}$, this state has the form:
\begin{eqnarray}
| \psi \rangle^{\text{ESC}}_{\text{Init}}  =  | \mathbf{x_0} \rangle_P \otimes \frac{1}{\sqrt{2}} \Big( e^{-i \beta_1} \cos \delta_1| L \rangle - e^{-i \alpha_1} \sin \delta_1 | D \rangle  \nonumber\\ 
 - e^{-i (\alpha_2 + \beta_1 - \beta_2)} \sin \delta_1 | U \rangle   -  e^{-i (\alpha_1 + \alpha_2 - \beta_2 - \varphi)} \cos \delta_1 | R \rangle \Big)\,. \nonumber\\
\label{escapingstate}
\end{eqnarray}
We note that the strong trapping effect can be avoided by starting the walk from an initially non-localized state. We studied the strength of the trapping effect numerically and illustrated it on Fig.~\ref{Fig_strtrap}.

The $\mathcal{C}$-class coins would make good candidates to trap and manipulate wave packets --- light pulses in the photonic implementation of experiments. The expected behavior of DTQW dynamics driven by a general coin is the propagation of the wavefunction over the lattice. However, changing the coin to a $\mathcal{C}$-class coin can trap (stop) a part of the spreading wave. This trapped pulse is held in its place by the trapping coin, and can be released by changing back to the original non-trapping coin.

\begin{figure}[t!]
\begin{center}
\includegraphics[width=0.48\textwidth]{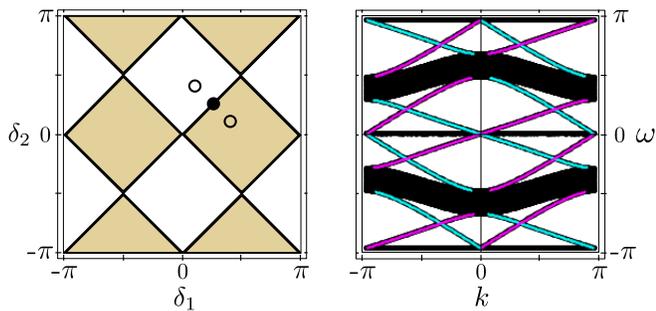}
\end{center}
\caption{ (Color online)
Topological map (left plot) and quasi-energy spectrum (right plot) of $\mathcal{C}$-class coins [see Eq.  (\ref{Cclass})].
The light brown and white areas correspond to different topological phases, i.e. in the boundary between two topologically different bulk regions we found a topologically protected edge state.
The thick lines represent the gap closing condition (\ref{condition_of_superlocalization}), which is also the condition describing the disappearance of strong trapping.
We used the definitions  (\ref{Cclass}) and (\ref{U2parametrisation}) with parameter values
$ \alpha_1 = \alpha_2 = \beta_1 = \beta_2 = \frac{\pi}{2}$ and $ \varphi = \pi$, invoking a real valued subclass of $\mathcal{C}$, which contains the Grover coin for $\delta_1 = \delta_2 = \frac{\pi}{4}$ (marked by the black dot).  The quasi-energy $\omega$ spectrum as the function of momentum $k$ on the right plot corresponds to the two-dimensional QW on a torus using two parallel bulk regions with $\delta_1 = \frac{\pi}{10}, \delta_2 = \frac{4\pi}{10}$ and $\delta_1 = \frac{4\pi}{10}, \delta_2 = \frac{\pi}{10}$ (denoted by circles on the topological map on the left). The topologically protected edge states are emphasized by magenta and cyan shading.
} \label{Figuretopo_spectrum}\end{figure}

The symmetries of the quasi-energy spectra allow for non-trivial topological phases in quantum walks \cite{Rudner2009,Kitagawa2010b,Asboth2012,Kitagawa2012,Asboth2013,Asboth2014,AsbothPreprint}. The controlled opening and closing of gaps in the quasi-energy spectrum can be a signature of the presence of such topological phases. We found that the spectra of weakly localizing walks satisfying Eq. (\ref{condition_of_superlocalization}) are gapless, whereas in the case of strong trapping, gaps open around the flat bands at $\omega = 0, \pi$ quasi-energies. Thus, parameters $\delta_1$ and $\delta_2$ tune the continuity of the spectrum. To map possible topological phases, we numerically searched this parameter space. We found that DTQWs with two contacted bulk regions using certain pairs of coins lead to the appearance of edge states. The found edge states are non-degenerate, can propagate only in one direction determined by the layout of the two bulk regions, and are also topologically protected from the decoherence caused by stationary (spatial) noise. The obtained topological map for the $\mathcal{C}$-class and a spectrum exhibiting edge states are shown in Fig.~\ref{Figuretopo_spectrum}. We note that up to our knowledge this is the first case when topological effects are reported in simple two-dimensional DTQWs using a single coin.

It is interesting to note that the $\mathcal{C}$-class walks can be linked with the so-called split-step walks. The concept of split-step (alternate step) two-dimensional quantum walks open a way to simulate some DTQWs using only a two-dimensional coin space \cite{Franco2011}. The time evolution of such walks is given as
\begin{equation}
 U_{\text{split}} = S_{\leftrightarrow} C_2 S_{\updownarrow} C_1\,,
 \label{splitstep2d_def}
\end{equation}
where $C_{k=\{1,2\}} \in SU(2)$. $S_{\updownarrow} (S_{\leftrightarrow})$ displaces the particle on a two-dimensional lattice down/up (left/right) with respect to its actual coin (qubit) state.  This construction implies that the particle cannot step backwards: Split-step walks live on the directed square lattice. Consequently, such walks can maintain their ballistic spreading better under noisy conditions  \cite{Chandrashekar2013}. The split-step QW also exhibits topological effects \cite{Kitagawa2010b,Asboth2013,Asboth2014,AsbothPreprint}. 

Let us rewrite the time evolution (\ref{splitstep2d_def}) of the split-step walk in the Fourier-picture, while relabeling the momenta. This results the following time evolution operator:
\begin{eqnarray}
\tilde{U}_{\text{split}}  & = &  \text{Diag}(e^{-i (k - l)/2}, e^{i (k - l)/2}) \cdot C_2 
\times \nonumber\\ && \text{Diag}(e^{-i (k + l)/2}, e^{i (k + l)/2}) \cdot C_1 \,.
\nonumber\\
\end{eqnarray}
Note that the relabeling of the momenta has the walker step in the diagonal and anti-diagonal directions on the lattice. An investigation of the spectrum of this model reveals a surprising fact: It is perfectly identical to the non flat band spectrum of the regular two-dimensional DTQW governed by the $\mathcal{C}$-class (\ref{Cclass}) with $\varphi = 0$.

This result unveils a correspondence between the two models, which helps us to draw consequences on the expected behavior. For example, the group velocities of the propagating peaks \cite{Peaks} and weak-limit distributions \cite{Grimmett2004,Watabe2008} should be similar. Moreover, the analytical tools proposed for the split-step models might be employed for $\mathcal{C}$-class based walks. The spectral correspondence also strengthens our observations on the topological effects.

Finally, let us briefly comment on the applicability of the new class of coins for the quantum search.
Two-dimensional DTQWs are suitable for searching marked elements on a $M \times M$ torus \cite{Search}. By design they allow for finding a single marked vertex in $O( \sqrt{N \log N})$ steps, with $M^2 = N$. The walk performing the search is a regular two-dimensional DTQW on the torus employing a transformed 4$\times$4 Grover coin $C^{(G)}(\sigma_x \otimes \sigma_x)$. In fact, this transformed coin is also in the $\mathcal{C}$-class with parameters $\delta_1 = \delta_2 = \frac{\pi}{4}$, $\alpha_1 = \beta_2= \frac{\pi}{2}$, $\alpha_2 = \beta_1= - \frac{\pi}{2}$ and  $\varphi = \pi$. At the marked vertex --- which the algorithm is designed to find --- the coin is changed to $- \sigma_x \otimes \sigma_x$. The initial state of the system is an equal superposition of all states in the natural basis, which is evolved for $\approx \sqrt{2N}$ steps. Following, a position measurement yields the marked vertex with probability $O( 1 / \log N)$. To realize a practically useful probability in order of $O( 1 )$, the algorithm should be repeated for $O( \sqrt{\log N})$ times and through amplitude amplification \cite{Grover1998} the $O( 1 )$ probability is achieved. In summary, the total runtime is $O( \sqrt{N \log N})$ steps.

\begin{figure}[t!]
\begin{center}
\includegraphics[width=0.45\textwidth]{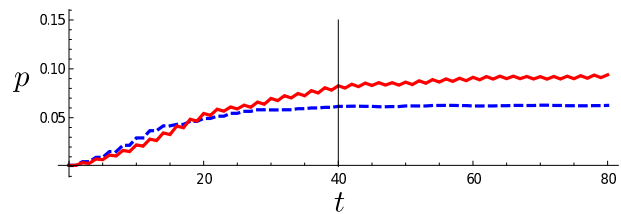} 
\end{center}
\caption{
(Color online)
Probability of finding a marked vertex on a $10 \times 80$ torus using the original spatial quantum search algorithm (blue dashed curve), and a modified one (red curve) employing the coin class $\mathcal{C}$ defined in Eq. (\ref{Cclass}) with parameters $\delta_1 = \delta_2 = \frac{31\pi}{90}$, $\alpha_1 = \beta_2= \frac{\pi}{2}$, $\alpha_2 = \beta_1= - \frac{\pi}{2}$ and  $\varphi = \pi$. The vertical black line denotes the measuring time defined in the algorithm $\sqrt{2LM} = 40$. At that time the probabilities of the original and modified algorithms at the marked vertex are $0.0616$ and  $0.0827$ respectively. Note that these probabilities are obtained before the final amplitude amplification step concluding the algorithm.
} \label{Figsearch}
\end{figure}

This algorithm is designed to perform searches on $M \times M$ tori, and on a general $L \times M$ torus it turns out to be suboptimal. However, employing the $\mathcal{C}$-class coins one can possibly optimize the algorithm for general tori. Our preliminary numerical studies show that for non-square tori a proper $\mathcal{C}$-class coin outperforms the original transformed Grover coin. See Fig.~\ref{Figsearch}. for details.

Finding coins from the $U(4)$ group for two-dimensional DTQWs that fulfill a specific goal is definitely a hard task. In this paper, we presented a coin class which exhibits a trapping feature similar to the Grover coin. The proposed many parameter, analytically given class is a considerable step towards classification of coins available on regular lattices. DTQWs driven by the new coin exhibit a novel phenomenon which we term as ``strong trapping", allowing for the trapping of all localized initial states. Generalizations to higher dimensional lattices are also possible allowing for constructing strongly trapping walks. The closed form of the coin matrix allows for further theoretical studies. We also demonstrated the topological properties of the class, revealing non-trivial topological phases, which were previously unobserved for simple single-coin two-dimensional DTQWs. The proposed system has a spectral connection with two-dimensional split-step quantum walks, giving a bridge between the two models. We have revealed that the quantum walk based search algorithm on a non-square torus benefits from the proposed coin, possibly allowing for optimization.

The proposed coin class is potentially very useful. These coins are promising choices for two-dimensional experiments \cite{2DExp}, as all of the necessary elements are currently available. Such experiments will allow to study topological effects, trapping, strong trapping, search, and other quantum walk related effects in a new setting.

We thank J. K. Asb\'oth and M. \v Stefa\v n\'ak for stimulating discussions.
We acknowledge support by GACR 13-33906S, RVO 68407700, the
Hungarian Scientific Research
Fund (OTKA) under Contract Nos. K83858, NN109651 and
the Hungarian Academy of Sciences (Lend\"ulet Program,
LP2011-016).

\end{document}